# Optimizing Clinical Trial Protocols Using EHR-Derived Heterogeneous Treatment Effects

Xiaodi Li, Ph.D.[1], Munhuwan Lee, Ph.D.[1], Pengyang Li, M.D., M.Sc.[2], Xiaoke Liu, M.D., Ph.D.[3], Jose K. James, M.D., Ph.D.[4], Patricia A. Pellikka, M.D.[5], Cui Tao, Ph.D.[1,*], and Nansu Zong, Ph.D.[1,*]

[1]Department of Artificial Intelligence and Informatics, Mayo Clinic, Rochester, MN, USA; [2]Division of Cardiology, Pauley Heart Center, Virginia Commonwealth University, Richmond, VA, USA; [3]MCHS Cardiology, Mayo Clinic, Rochester, MN, USA; [4]Mayo Clinic School of Graduate Medical Education, Mayo Clinic, Rochester, MN, USA; [5]Department of Cardiovascular Medicine, Mayo Clinic, Rochester, MN, USA

* Corresponding authors. Email: Tao.Cui@mayo.edu; Zong.Nansu@mayo.edu

## Abstract

Traditional randomized trials often obscure clinically meaningful heterogeneity in treatment response by focusing on average effects. Leveraging real-world data to emulate clinical trials and estimate heterogeneous treatment effects (HTEs) offers a promising path toward more precise and efficient trial design. In this study, we emulate the DAPA-HF trial using electronic health records from the Mayo Clinic Cloud (MCC) to investigate whether HTE-guided stratification can identify patient subgroups with distinct treatment responses to dapagliflozin versus placebo in patients with heart failure with reduced ejection fraction. All-cause mortality was evaluated using Cox proportional hazards models, with HTEs estimated using a Meta-S learner and subgroups defined using a decision tree–based thresholding approach. In the overall cohort of the emulation, no significant treatment difference was observed (HR, 1.681; 95% CI, 0.828–3.413; p = 0.1507). However, compared with the overall emulated cohort, in which dapagliflozin showed no statistically significant survival benefit, HTE-driven stratification identified subgroups with significant and directionally distinct treatment effects. The beneficial (low-HTE) subgroup showed a significant survival benefit from dapagliflozin (HR = 0.203, 95% CI, 0.087–0.476, p = 0.0002), whereas the harmful (high-HTE) subgroup showed a significant harmful association with markedly increased mortality risk (HR = 6.680, 95% CI, 2.759–16.171, p < 0.0001). These findings indicate that HTE-guided stratification can uncover clinically meaningful beneficial and harmful treatment-effect patterns that are masked in the full-cohort emulation. These results demonstrate that HTE-guided trial emulation can uncover clinically actionable subpopulations hidden within aggregate analyses, highlighting a powerful framework for advancing precision medicine and next-generation clinical trial design.

**Keywords:** Heterogeneous Treatment Effect; RCT emulation; real-world data; electronic health records; cardiovascular treatment; observational study

## 1. Introduction

Randomized controlled trials (RCTs) remain the gold standard for establishing causal evidence in clinical research and guiding treatment guidelines. However, even rigorously designed large-scale trials may yield neutral or inconclusive findings, partly because average treatment effects can obscure heterogeneous responses across patient subgroups, raising challenges for trial efficiency, patient selection, and real-world generalizability [1, 2]. One of the reasons for such failures lies in protocol design: eligibility criteria and treatment strategies are often determined based on expert consensus rather than data-driven insights. Protocols that are too restrictive limit external validity, while overly broad inclusion criteria may dilute treatment effects across heterogeneous patient populations, leading to statistically insignificant outcomes [3-5].

Recent studies on eligibility criteria relaxation underscore the importance of balancing inclusiveness with statistical efficiency. Liu et al. introduced Trial Pathfinder, a framework that systematically relaxes exclusion rules in oncology trials using real-world data, showing that broader criteria can double eligible populations while maintaining safety signals [6]. Similarly, Kaur et al. demonstrated that relaxing organ function and performance status thresholds improved diversity and inclusivity across 22 cancer types, though with modest trade-offs in survival outcomes [7]. Sen et al. developed quantitative indices (e.g., GIST 2.0) to measure representativeness of trial-eligible populations [8]. Collectively, these works suggest that data-driven eligibility design is critical for trial optimization, ensuring both representativeness and statistical power. In response, the biomedical research community has increasingly emphasized trial optimization strategies that leverage real-world data and advanced analytics to optimize study design, improve statistical power, and enhance subgroup identification [9-11].

Within this context, Heterogeneous Treatment Effect (HTE) estimation has emerged as a powerful methodological tool. The estimation of heterogeneous treatment effects (HTEs) is central to precision medicine, as it enables therapeutic decisions to be tailored according to individual patient characteristics and expected treatment responsiveness. Randomized controlled trials (RCTs) remain the benchmark for evaluating average treatment effects (ATEs); however, their primary focus on population-level effects may obscure clinically important variation in treatment response across patient subgroups [12, 13]. In addition, the high cost, restrictive eligibility criteria, and long timelines of conventional RCTs limit their scalability and may reduce their generalizability to diverse clinical populations [14, 15]. These limitations highlight the need for scalable approaches that can complement RCTs by supporting individualized treatment-effect inference in real-world settings.

These considerations motivate the need to move beyond population-average treatment effects and explicitly examine treatment-effect heterogeneity. Trial emulation primarily estimates the marginal treatment effect in the overall emulated cohort, commonly expressed as the Average Treatment Effect (ATE). This estimate is useful for evaluating whether a treatment shows benefit at the population level, but it does not explain whether different patient subgroups respond differently. In contrast, HTE analysis estimates conditional treatment effects, such as Conditional Average Treatment Effects (CATEs) or Individual Treatment Effects (ITEs), to assess how treatment benefit or risk varies across patient characteristics. This distinction is important because a treatment may show no statistically significant effect in the full cohort emulation, yet still demonstrate significant benefit in one subgroup and significant harm in another.

Numerous studies have demonstrated that trial emulation can reproduce both effectiveness and safety endpoints with high fidelity to prospective trials [9, 15]. At the same time, machine learning approaches for HTE estimation, such as causal forests, meta-learners, and representation learning, have been increasingly applied to estimate individualized treatment effects in complex observational data [16-18]. Despite these advances, an important gap remains: HTE estimates are often reported as model-derived scores without being translated into clinically interpretable patient stratification. As a result, it remains difficult to determine how estimated heterogeneity should inform trial design, eligibility optimization, or subgroup-specific treatment decisions. Moreover, many subgroup analyses still rely on heuristic or post hoc definitions rather than outcome-based evidence of significant treatment benefit or harm.

To address this gap, we introduce a data-driven framework that uses HTE estimates not merely as predictive scores, but as a bridge between population-level trial emulation and clinically actionable patient stratification. The central goal of this framework is to determine whether an apparently neutral or modest treatment effect in the overall emulated cohort may conceal subgroups with significant benefit or potential harm. To achieve this, individualized HTE estimates are used to guide the identification of response-based patient subgroups, rather than relying on predefined or post hoc subgroup definitions. We then translate these HTE-derived patterns into interpretable strata through decision tree–based thresholding, allowing treatment-effect heterogeneity to be expressed as clinically meaningful cohort definitions. Finally, survival-based validation is performed within the resulting subgroups to determine whether the HTE-guided stratification corresponds to statistically and clinically meaningful differences in treatment outcomes. In this way, the proposed framework provides a mechanism for moving from average treatment-effect estimation toward precision cohorting and eligibility optimization in real-world trial emulation.

We selected the DAPA-HF (Dapagliflozin and Prevention of Adverse Outcomes in Heart Failure, NCT03036124) trial as the target trial for emulation because it provides a well-characterized benchmark for evaluating dapagliflozin in patients with heart failure with reduced ejection fraction (HFrEF) [19]. DAPA-HF was a multicenter, randomized, double-blind, placebo-controlled trial that enrolled 4,744 patients with HFrEF.

Although its primary endpoint was a composite cardiovascular outcome, the trial also evaluated all-cause mortality as a secondary outcome over a median follow-up of 27.8 months, providing a clinically meaningful survival endpoint for real-world emulation. In the original trial, dapagliflozin was associated with a significant reduction in all-cause mortality, with 276 deaths (11.6%) in the dapagliflozin group compared with 329 deaths (13.9%) in the placebo group. This corresponded to a hazard ratio of 0.83 (95% CI, 0.71–0.97; p = 0.0217) based on a Cox proportional hazards model stratified by type 2 diabetes status, establishing an important reference effect against which the emulated full-cohort and HTE-stratified results can be interpreted.

This HTE-based trial emulation optimization framework offers four key contributions. (1) Although our DAPA-HF emulation shows no statistically significant treatment effect at the population level, our approach uncovers directionally distinct treatment effects across HTE-defined subgroups, including both strongly beneficial and potentially harmful responses, thereby revealing clinically meaningful heterogeneity that is obscured in aggregate analyses. (2) We move beyond conventional use of HTE as a latent model output by introducing HTE-based subgrouping strategies that transform continuous treatment-effect estimates into clinically interpretable patient strata, enabling transparent identification of responder and non-responder populations. (3) We establish a validation pipeline that links HTE-based stratification to downstream survival analysis, demonstrating that the identified subgroups correspond to statistically significant and clinically meaningful differences in mortality outcomes. (4) We further translated HTE-derived subgroup findings into interpretable eligibility-criteria optimization by using decision-tree–based rule discovery to generate optimized clinical rule paths, which were then incorporated into cohort reconstruction and re-evaluated through the full trial emulation and Cox survival-analysis workflow. By integrating real-world trial emulation, HTE estimation, survival validation, and eligibility criteria optimization, our framework reframes trial emulation from estimating average treatment effects to identifying actionable patient subpopulations and optimizing future trial design.

## 2. Results

### *2.1.Study Details*

Table 1 summarizes key characteristics from the DAPA-HF trial alongside corresponding patient counts from the Mayo Clinic Cloud (MCC) cohort. It also presents cohort sizes under different stratification strategies, providing a comparative view of population distributions across the original trial and the real-world emulated data. The setting is shown with patient distributions stratified by treatment group, sex, and race. While the original trial data is limited in some demographic breakdowns, the MCC cohort offers more complete representation across subgroups. This comparison highlights the alignment and divergence between trial-based and clinical populations, offering insight into generalizability and potential for heterogeneity analysis.

**Table 1.** Baseline Demographic and Clinical Characteristics of the MCC Real-World Cohort and the Original DAPA-HF Trial. This table illustrates the demographic and clinical distributions of patients across subgroups for both the real-world MCC cohort and the original DAPA-HF trial.

| Subgroup | MCC | DAPA-HF Trial |
|---|---|---|
| Treatment: Dapagliflozin | 149 (1.65%) | 2,373 (50.02%) |
| Treatment: Placebo | 8,865 (98.35%) | 2,371 (49.98%) |
| Sex: Male | 5,851 (64.91%) | 3,635 (76.62%) |
| Sex: Female | 3,163 (35.09%) | 1,109 (23.38%) |
| Race: White | 8,573 (95.10%) | 3,333 (70.26%) |
| Race: Black | 164 (1.82%) | 226 (4.76%) |
| Race: Other | 102 (1.13%) | 1,185 (24.98%) |
| Race: Unknown | 175 (1.94%) | 0 (0.00%) |

| Ethnicity: Hispanic or Latino | 115 (1.28%) | N/A |
|---|---|---|
| Enrollment eGFR ≥ 30 | 9,014 (100.00%) | N/A |
| Previous SGLT2 intolerance | 104 (1.15%) | N/A |
| Severe renal disease at randomization | 3,623 (40.19%) | N/A |
| Recent myocardial infarction | 3,475 (38.55%) | N/A |
| Recent unstable angina | 1,324 (14.69%) | N/A |
| Recent stroke | 2,709 (30.05%) | N/A |
| Recent TIA | 881 (9.77%) | N/A |
| Recent PCI | 2,634 (29.22%) | N/A |
| Recent CABG | 1,560 (17.31%) | N/A |
| Recent valvular procedure | 1,031 (11.44%) | N/A |
| Recent CRT implantation | 3,487 (38.68%) | N/A |
| Restrictive cardiomyopathy | 148 (1.64%) | N/A |
| Constrictive pericarditis | 35 (0.39%) | N/A |
| Hypertrophic obstructive cardiomyopathy | 110 (1.22%) | N/A |
| Primary valvular disease | 6,905 (76.60%) | N/A |
| Advanced heart block without pacemaker | 3,490 (38.72%) | N/A |
| HF duration ≥ 2 months | 1,216 (13.49%) | N/A |
| Elevated NT-proBNP | 9,014 (100.00%) | N/A |
| Baseline LVEF subgroup | 9,014 (100.00%) | N/A |
| Baseline NT-proBNP subgroup | 9,014 (100.00%) | N/A |
| MRA at baseline | 3,213 (35.64%) | N/A |
| Type 2 diabetes at baseline | 4,197 (46.56%) | N/A |
| Atrial fibrillation/flutter on enrollment ECG | 6,309 (69.99%) | N/A |
| Main HF cause subgroup: Ischemic | 7,794 (86.47%) | N/A |
| Baseline BMI subgroup | 9,014 (100.00%) | N/A |
| Dapagliflozin dose subgroup | 9,014 (100.00%) | N/A |

## *2.2. Results*

### 2.2.1. Goal and HTE-based Stratification Design

The HTEs estimated by meta-learning algorithms allow us to model individual variability in response to treatment. To translate these continuous HTE estimates into clinically meaningful subgroups, we employed two complementary stratification strategies in parallel. First, we applied a decision tree–based thresholding approach, in which a shallow regression tree was trained on the transformed outcome, incorporating both mortality events and treatment assignment, to learn a data-driven HTE cutoff that separates patients with different treatment-response profiles.

### 2.2.2. Threshold (Quantile) Selection and Final Cohort Definition

To define clinically meaningful HTE-based subgroups, we first employed a decision tree–based thresholding approach to derive data-driven cutoffs. The final threshold is -0.03501 (HTE value). The results are summarized in Table 2. Using the learned threshold, the original cohort can be stratified into subgroups predicted to benefit from treatment (HR = 0.203, p = 0.0002) and those predicted to experience potential harm (HR = 6.680, p < 0.0001), with statistically significant differences observed between these groups. This approach provides an interpretable and objective criterion for partitioning patients based on predicted treatment effects.

**Table 2.** Decision Tree–based Thresholding Results for Meta-S Learners (threshold = -0.03501).

| Model | Meta-S | | | | | | |
|---|---|---|---|---|---|---|---|
| HTE Group = 0 (HTE < threshold) | | | | HTE Group = 1 (HTE > threshold) | | | |
| Beneficial | | | | Harmful | | | |
| HR | p values | Placebo | Dapagliflozin | HR | p values | Placebo | Dapagliflozin |
| 0.203 | 0.0002 | 3561 | 47 | 6.680 | < 0.0001 | 5304 | 102 |

### 2.2.3. Survival Curve Evidence of Treatment Effect Heterogeneity

As shown in Figure 1, the survival curves illustrate how different HTE-based stratification strategies influence the observed treatment effects between placebo and dapagliflozin. In the unstratified (standard emulation) cohort, the survival curves were closely aligned, indicating no statistically significant difference in mortality between treatment groups. In contrast, the decision tree–based thresholding approach produced subgroups in which the separation between treatment arms became more pronounced. In particular, subgroups predicted to benefit from treatment consistently showed improved survival under dapagliflozin, whereas other subgroups exhibited attenuated or negligible effects. Notably, the consistency of this pattern across two distinct HTE estimation strategies supports the robustness of the observed heterogeneity. Overall, these results demonstrate that HTE-based stratification can reveal clinically meaningful differences in treatment response that remain obscured in population-level analyses.

(a) Standard Emulation

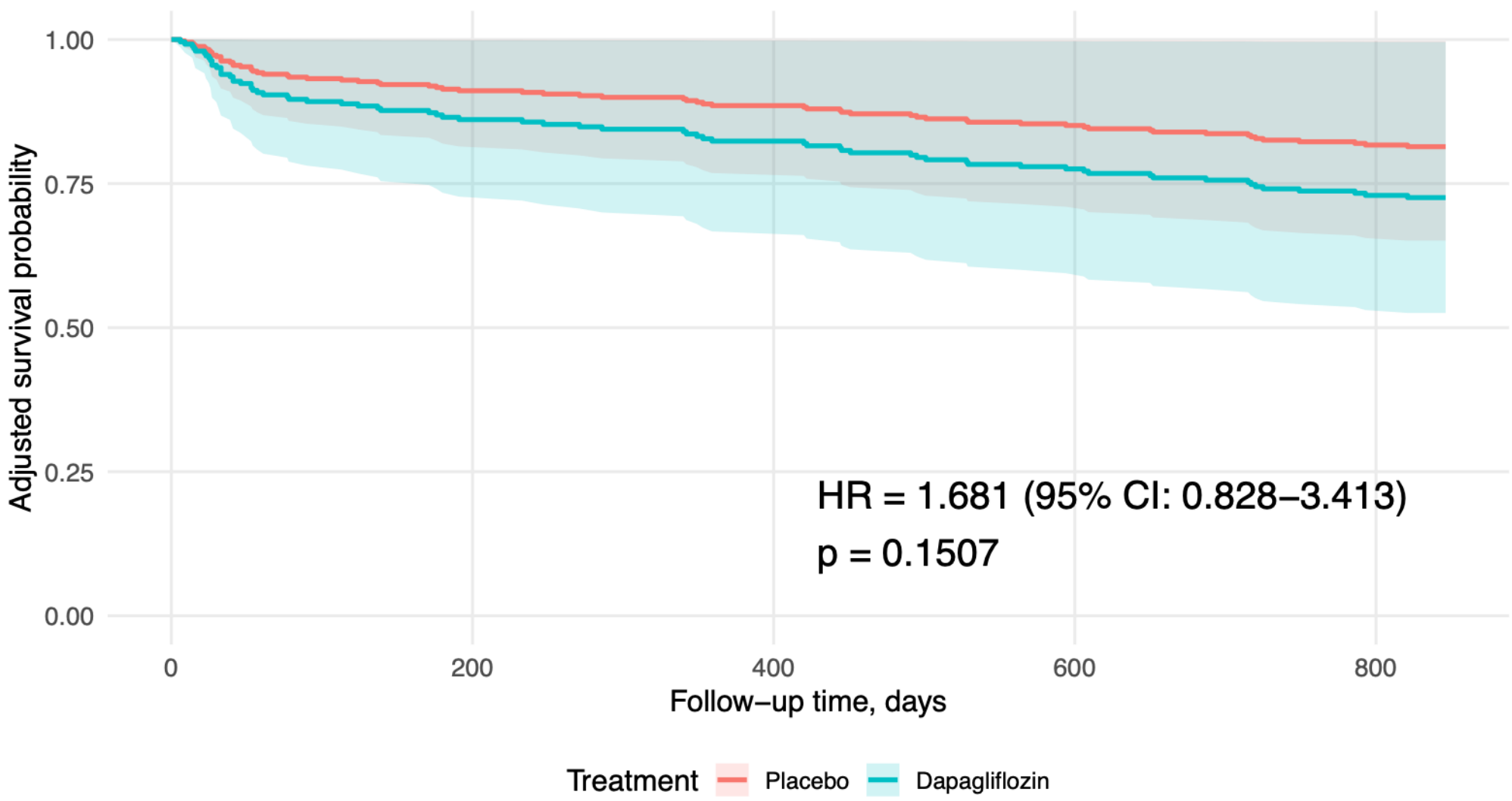

(b) Decision Tree-based Thresholding, Beneficial Group

Adjusted Survival Curves for Mortality: Propensity–Matched Cohort

Counterfactual standardized survival from existing adjusted Cox model

Adjusted survival probability

1.00

0.75

0.50

0.25

0.00

0 200 400 600 800

Follow–up time, days

HR = 0.203 (95% CI: 0.087–0.476)

p = 0.0002

Treatment Placebo Dapagliflozin

(c) Decision Tree-based Thresholding, Harmful Group

Adjusted Survival Curves for Mortality: Propensity–Matched Cohort

Counterfactual standardized survival from existing adjusted Cox model

Adjusted survival probability

1.00

0.75

0.50

0.25

0.00

0 200 400 600 800

Follow–up time, days

HR = 6.680 (95% CI: 2.759–16.171)

$p < 0.0001$

Treatment Placebo Dapagliflozin

**Figure 1.** Survival Curves Stratified by HTE-based Decision Tree-based Thresholding. The panel (a) shows the standard emulation results, where no statistically significant difference in survival is observed between the dapagliflozin (blue) and placebo (red) groups. The panels (b) and (c) present survival curves stratified by HTE-derived subgroups (low and high HTE) using a decision tree–based threshold. In the beneficial subgroup, a clearer separation between treatment arms emerges, indicating improved survival among patients predicted to benefit from dapagliflozin. In contrast, the harmful subgroup shows attenuated or reversed treatment effects. Overall, HTE-based stratification reduces overlap between survival curves and reveals heterogeneous treatment effects that are not apparent in the population-level analysis.

2.2.4. Hazard Ratio Quantification and Covariate Effects

As shown in Figures 2 and 3, HTE-based stratification revealed substantial heterogeneity in the treatment effect of dapagliflozin on mortality. In the overall propensity-matched cohort, the treatment effect was not statistically significant (HR, 1.681; 95% CI, 0.828–3.413; $p = 0.1507$), indicating no clear difference between treatment arms at the population level.

In contrast, the beneficial subgroup in Figure 2 demonstrates a strong protective association for dapagliflozin (HR = 0.203, 95% CI, 0.087–0.476, $p = 0.0002$). Within this subgroup, several covariates are significantly associated with mortality. Increased mortality risk is observed for Recent PCI (HR = 12.428, 95% CI, 4.547–33.972, $p < 0.0001$), Advanced heart block without pacemaker (HR = 5.668, 95% CI, 2.283–14.071, $p = 0.0002$), HF duration >= 2 months (HR = 23.139, 95% CI, 2.963–180.681, $p = 0.0027$), Baseline NT−proBNP subgroup (HR = 1.000, 95% CI, 1.000–1.000, $p = 0.0003$), and Dapagliflozin dose subgroup (HR = 1.001, 95% CI, 1.001–1.002, $p = 0.0015$). On the contrary, several variables are associated with reduced mortality risk, including Female Sex (HR = 0.117, 95% CI, 0.027–0.507, $p = 0.0041$), Enrollment eGFR >= 30 (HR = 0.888, 95% CI, 0.872–0.904, $p < 0.0001$), Severe renal disease at randomization (HR = 0.005, 95% CI, 0.002–0.012, $p < 0.0001$), Recent unstable angina (HR = 0.004, 95% CI, 0.001–0.031, $p < 0.0001$), Recent valvular procedure (HR = 0.007, 95% CI, 0.001–0.050, $p < 0.0001$), Restrictive cardiomyopathy (HR = 0.207, 95% CI, 0.047–0.904, $p = 0.0362$), Baseline LVEF subgroup (HR = 0.998, 95% CI, 0.996–1.000, $p = 0.0171$), MRA at baseline (HR = 0.425, 95% CI, 0.183–0.990, $p = 0.0475$), and Baseline BMI subgroup (HR = 0.997, 95% CI, 0.996–0.998, $p < 0.0001$). Overall, these results suggest that dapagliflozin remains strongly associated with lower mortality in the beneficial subgroup, while mortality risk is also shaped by cardiovascular comorbidities, disease duration, renal function, and baseline clinical characteristics.

In the harmful subgroup (Figure 3), the direction of the treatment association is reversed, with dapagliflozin associated with increased mortality risk (HR = 6.680, 95% CI, 2.759–16.171, $p < 0.0001$). Several covariates are also significantly associated with higher mortality risk in this subgroup, including Severe renal disease at randomization (HR = 2.633, 95% CI, 1.178–5.885, $p = 0.0183$), Recent myocardial infarction (HR = 2.394, 95% CI, 1.094–5.240, $p = 0.0289$), Elevated NT−proBNP (HR = 1.000, 95% CI, 1.000–1.000, $p < 0.0001$), and Baseline LVEF subgroup (HR = 1.001, 95% CI, 1.000–1.002, $p = 0.0436$). In contrast, Female Sex is associated with reduced mortality risk (HR = 0.230, 95% CI, 0.064–0.823, $p = 0.0239$). Overall, these findings suggest that the harmful subgroup represents a clinically distinct population in which dapagliflozin is associated with higher mortality, with risk further shaped by renal disease, recent myocardial infarction, and baseline cardiac biomarkers.

Taken together, these results show that HTE-based stratification identifies clinically meaningful and directionally distinct treatment associations that are not evident in the unstratified analysis, highlighting substantial heterogeneity in treatment response across patient subgroups.

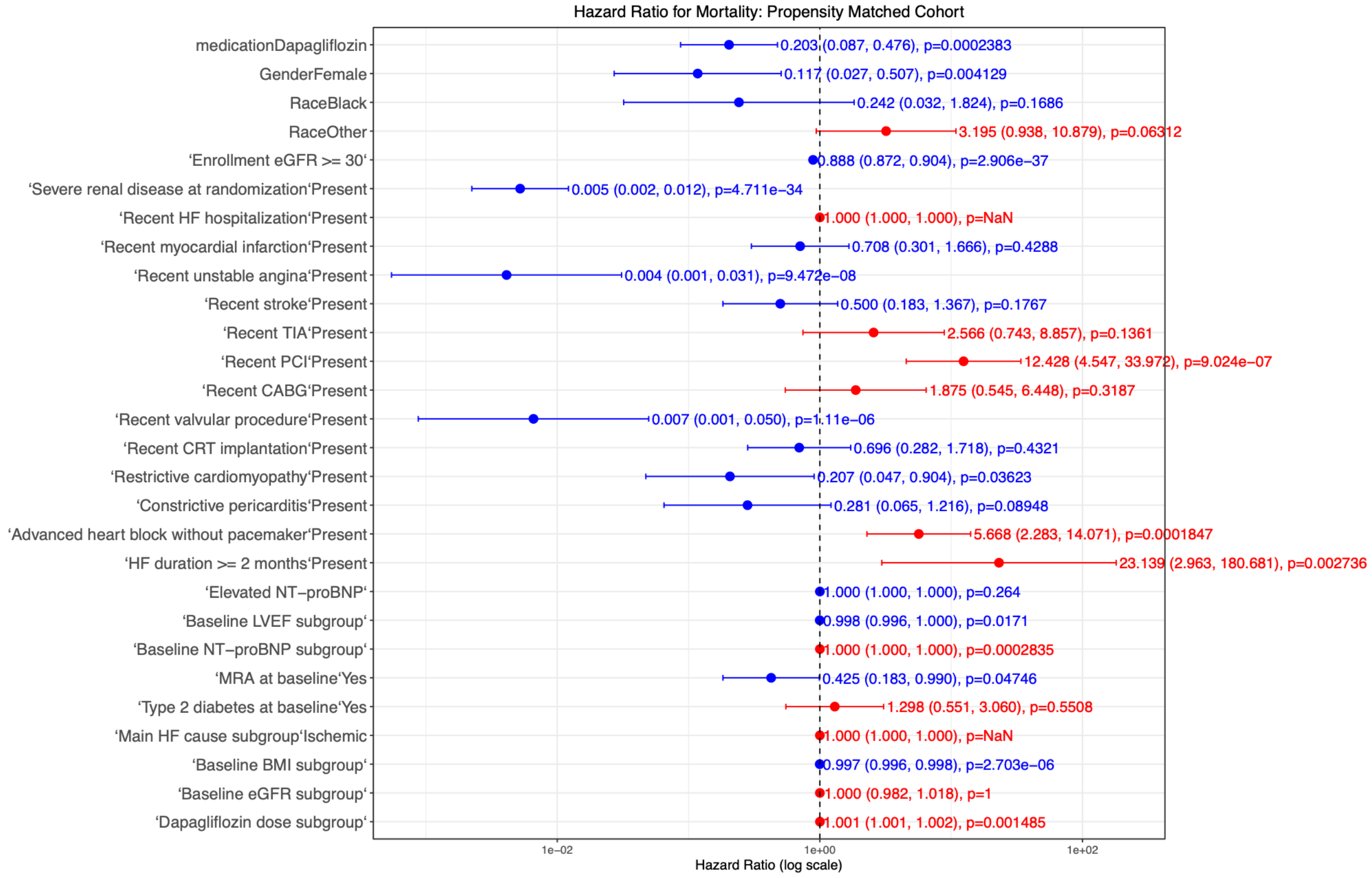

**Figure 2.** Hazard Ratios for Mortality in the Beneficial Subgroup from Decision Tree-based Thresholding. This figure presents hazard ratio estimates and 95% confidence intervals for dapagliflozin versus placebo and associated clinical covariates within the beneficial subgroup, defined as patients predicted to benefit from treatment. Estimates are obtained from a Cox proportional hazards model in the propensity-matched cohort. The hazard ratio for dapagliflozin versus placebo is shown in the top row (HR = 0.203), indicating a substantial reduction in mortality risk within this subgroup. Hazard ratios are displayed on a logarithmic scale, with horizontal lines representing confidence intervals; values below 1 indicate reduced mortality risk, whereas values above 1 indicate increased risk. The results highlight both the protective effect of dapagliflozin and the contribution of key clinical covariates within the subgroup identified by HTE-based stratification. Red markers denote increased risk (HR > 1), while blue markers represent protective effects (HR < 1). Annotated p-values highlight the strength of association for each factor.

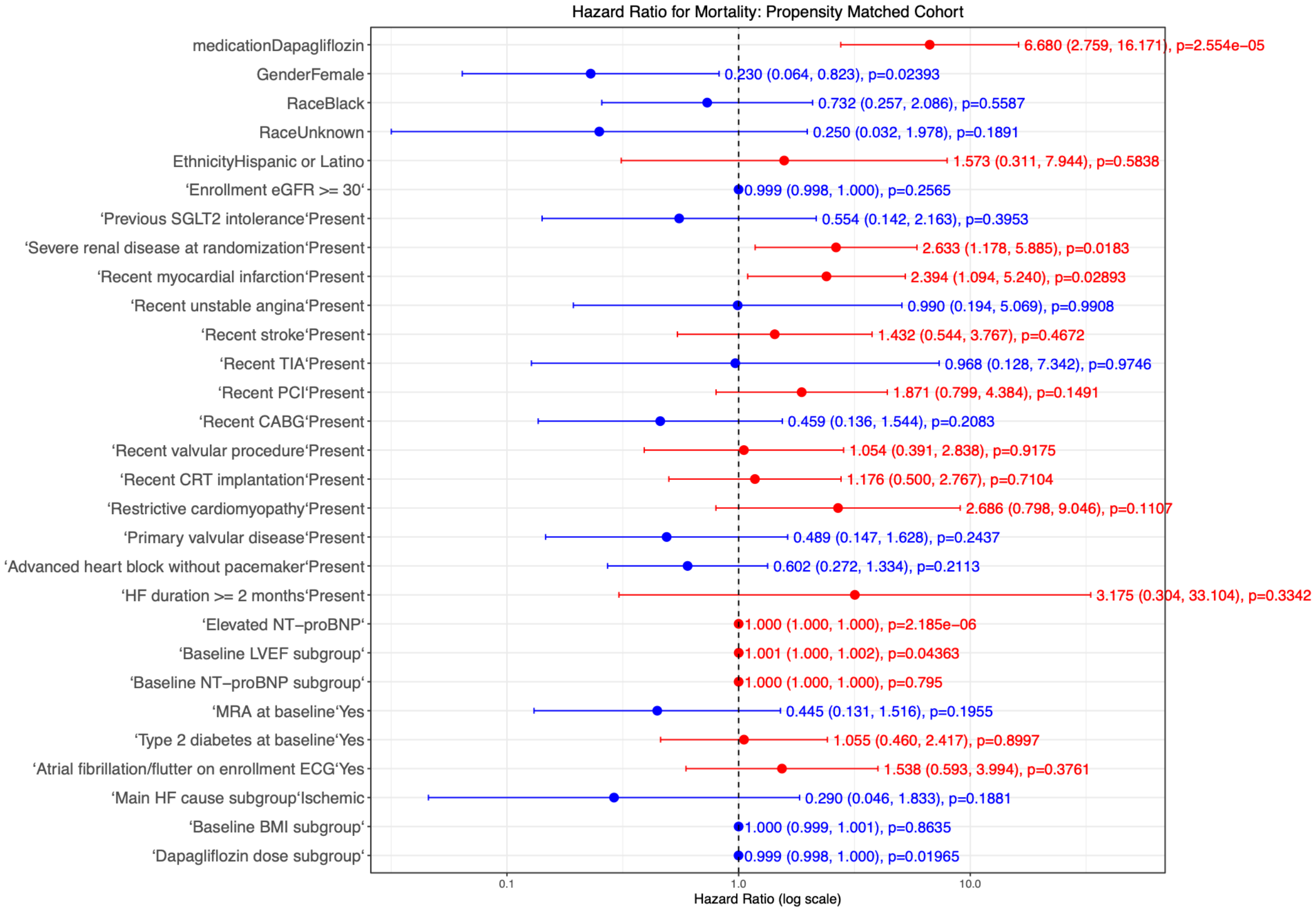

**Figure 3.** Hazard Ratios for Mortality in the Harmful Subgroup from Decision Tree-based Thresholding. This figure presents hazard ratio estimates and 95% confidence intervals for dapagliflozin versus placebo and associated clinical covariates within the harmful subgroup, defined as patients predicted to derive limited or no benefit from treatment. Estimates are obtained from a Cox proportional hazards model in the propensity-matched cohort. The hazard ratio for dapagliflozin versus placebo is shown in the top row (HR = 6.680), indicating a substantially increased mortality risk in this subgroup. Hazard ratios are displayed on a logarithmic scale, with horizontal lines representing confidence intervals; values below 1 indicate reduced mortality risk, whereas values above 1 indicate increased risk. Compared with the beneficial subgroup, the treatment effect is diminished, and the pattern of covariate associations differs, highlighting heterogeneity in treatment response across patient subgroups. Red markers denote increased risk ($HR > 1$), while blue markers represent protective effects ($HR < 1$). Annotated p-values highlight the strength of association for each factor.

### 2.2.5. Eligibility Criteria Optimization Based on HTE-Informed Decision-Tree

To translate the HTE-derived subgroup findings into interpretable eligibility-criteria optimization, we performed an HTE-informed rule-discovery and cohort-reconstruction analysis. First, patients were ranked according to their predicted HTE values, with lower HTE values representing potential treatment benefit and higher HTE values representing potential harm. A trial-referenced library of eligibility-related criteria and clinically meaningful baseline subgroup factors was then used to identify rule-based clinical paths that approximated these HTE-enriched benefit and harm targets. Multiple positive tree-derived paths were combined using OR logic to generate candidate optimized cohorts. These candidates were ranked and filtered using overlap-based metrics, including precision, recall, and Jaccard index, as well as cohort size, treatment/control balance, and event availability. Candidate cohorts passing these criteria were subsequently evaluated using propensity-score matching and adjusted Cox survival analysis. The final optimized beneficial and harmful rules were then incorporated into the cohort-construction step, and the full emulation workflow was repeated to obtain the final survival estimates.

As shown in Table 3, the full-emulation framework retained the core trial-defining inclusion criteria, including adult patients with symptomatic HFrEF, LVEF ≤ 40%, and eGFR ≥ 30 mL/min/1.73 m², while using a broad set of demographics, renal, cardiovascular, medication, and clinical subgroup variables as adjustment covariates. In the optimized emulation framework, the core eligibility criteria remained as strict inclusion criteria, but selected eligibility-related factors were reclassified into subgroup-defining rules for the beneficial and harmful cohorts. Because the optimized cohorts were generated as OR-unions of multiple tree-derived rule paths, “Strict Inclusion” and “Strict Exclusion” indicate that the criterion was used as a strict component in at least one selected optimized rule path; these labels should not be interpreted as criteria required uniformly across all patients in the optimized cohort. For the optimized beneficial cohort, several cardiovascular exclusion-related criteria, including no recent unstable angina, myocardial infarction, PCI, CABG, valvular procedure, CRT implantation, stroke, restrictive cardiomyopathy, and constrictive pericarditis were promoted to strict exclusion components, while type 2 diabetes at baseline and absence of MRA use were incorporated as subgroup-defining factors. In contrast, the optimized harmful cohort was primarily characterized by MRA use, atrial fibrillation or flutter on enrollment ECG, and absence of recent unstable angina as selected rule components. Criteria not selected in the final optimized rule but retained in the propensity-score or Cox adjustment models were classified as confounders, whereas non-operational or unavailable criteria remained classified as drop/operational. Overall, this optimization refines the original broad trial-emulation framework into two HTE-guided, clinically interpretable eligibility structures that distinguish patients more likely to benefit from dapagliflozin from those showing potential harm while preserving adjustment for remaining baseline differences.

The selected HTE-informed optimized eligibility-criteria subgroups showed distinct treatment-associated mortality patterns. In the beneficial subgroup, dapagliflozin was associated with substantially lower mortality risk compared with placebo, with an adjusted HR of 0.175 (95% CI, 0.047–0.650; $p = 0.0092$). In contrast, in the harmful subgroup, dapagliflozin was associated with increased mortality risk, with an adjusted HR of 2.125 (95% CI, 1.008–4.479; $p = 0.0476$). These findings suggest that the optimized eligibility-criteria framework identified clinically distinct subgroups with directionally opposite treatment associations.

Figure 4 presents the adjusted survival curves for the optimized eligibility-criteria beneficial and harmful subgroups in the propensity-matched cohorts. In the beneficial subgroup, the dapagliflozin curve remained consistently above the placebo curve during follow-up, indicating improved adjusted survival among treated patients. In the harmful subgroup, the dapagliflozin curve declined more rapidly and remained below the placebo curve, indicating worse adjusted survival among treated patients. The shaded areas represent 95% confidence intervals, reflecting uncertainty around the counterfactual standardized survival estimates.

Figures 5 and 6 present the covariate-adjusted hazard ratios for mortality in the propensity-matched beneficial and harmful cohorts, respectively. In the beneficial subgroup, dapagliflozin remained strongly associated with reduced mortality after adjustment, whereas several covariates showed wide confidence intervals, suggesting limited precision for some subgroup-specific estimates. In the harmful subgroup, dapagliflozin was associated with increased mortality risk after adjustment, while baseline clinical factors such as recent myocardial infarction and heart failure duration also showed elevated mortality associations. Overall, these forest plots support the survival-curve findings and illustrate that the optimized eligibility-criteria subgroups retained distinct treatment-risk profiles after covariate adjustment.

**Table 3.** Comparison of Full-Emulation Eligibility Criteria and Optimized Eligibility Criteria.

| **Original Trial** | **Full Emulation** | **Optimized Emulation (Beneficial)** | **Optimized Emulation (Harmful)** |
|---|---|---|---|
| **Inclusion Criteria** | **Full Emulation Criteria Classification** | **Optimized Criteria Classification** | **Optimized Criteria Classification** |
| Male or female aged ≥ 18 years | Strict Inclusion | Strict Inclusion | Strict Inclusion |
| Established documented diagnosis of symptomatic HFrEF (NYHA class II-IV) | Strict Inclusion | Strict Inclusion | Strict Inclusion |
| Left ventricular ejection fraction (LVEF) ≤ 40% | Strict Inclusion | Strict Inclusion | Strict Inclusion |

| eGFR ≥ 30 mL/min/1.73 $m^2$ by CKD-EPI at enrolment (visit 1) | Strict Inclusion | Strict Inclusion | Strict Inclusion |
|---|---|---|---|
| Heart failure present for at least 2 months | Confounder | Confounder | Confounder |
| Elevated NT-proBNP levels | Confounder | Confounder | Confounder |
| Provision of signed informed consent prior to any study-specific procedures | Not included | Not included | Not included |
| Receiving background standard of care for HFrEF according to locally recognized guidelines | Not included | Not included | Not included |
| **Exclusion Criteria** | **Full Emulation Criteria Classification** | **Optimized Criteria Classification** | **Optimized Criteria Classification** |
| Unstable angina within 12 weeks prior to enrolment | Confounder | Strict Exclusion | Strict Exclusion |
| Heart failure due to restrictive cardiomyopathy | Confounder | Strict Exclusion | Confounder |
| Stroke within 12 weeks prior to enrolment | Confounder | Strict Exclusion | Confounder |
| Heart failure due to constrictive pericarditis | Confounder | Strict Exclusion | Confounder |
| Percutaneous coronary intervention within 12 weeks prior to enrolment | Confounder | Strict Exclusion | Confounder |
| Myocardial infarction within 12 weeks prior to enrolment | Confounder | Strict Exclusion | Confounder |
| Coronary artery bypass grafting within 12 weeks prior to enrolment | Confounder | Strict Exclusion | Confounder |
| Valvular repair or replacement within 12 weeks prior to enrolment | Confounder | Strict Exclusion | Confounder |
| Cardiac resynchronization therapy (CRT) implantation within 12 weeks prior to enrolment | Confounder | Strict Exclusion | Confounder |
| Planned percutaneous coronary intervention after randomization | Not included | Strict Exclusion | Not included |
| Planned coronary artery bypass grafting after randomization | Not included | Strict Exclusion | Not included |
| Planned valvular repair or replacement after randomization | Not included | Strict Exclusion | Not included |
| Intent to implant a CRT device | Not included | Strict Exclusion | Not included |
| Heart failure due to hypertrophic obstructive cardiomyopathy | Confounder | Confounder | Confounder |
| Previous intolerance of an SGLT2 inhibitor | Strict Exclusion | Confounder | Confounder |
| Severe renal disease: eGFR $<30$ mL/min/1.73 $m^2$ by CKD-EPI at time of randomization | Strict Exclusion | Confounder | Confounder |
| Unstable renal disease at time of randomization | Not included | Confounder | Confounder |
| Expected implantation of a ventricular assist or similar device after randomization | Strict Exclusion | Confounder | Confounder |
| Type 1 diabetes mellitus | Strict Exclusion | Confounder | Confounder |
| Symptomatic hypotension | Strict Exclusion | Confounder | Confounder |

| Systolic blood pressure <95 mmHg at 2 of 3 measurements at either visit 1 or visit 2 | Strict Exclusion | Confounder | Confounder |
|---|---|---|---|
| Previous cardiac transplantation | Strict Exclusion | Confounder | Confounder |
| Previous implantation of a ventricular assist device or similar device | Strict Exclusion | Confounder | Confounder |
| Hospitalization due to decompensated heart failure within 4 weeks prior to enrolment | Confounder | Confounder | Confounder |
| Transient ischemic attack within 12 weeks prior to enrolment | Confounder | Confounder | Confounder |
| Heart failure due to uncorrected primary valvular disease | Confounder | Confounder | Confounder |
| Second- or third-degree heart block without a pacemaker | Confounder | Confounder | Confounder |
| Receiving therapy with an SGLT2 inhibitor within 8 weeks prior to enrolment | Not included | Not included | Not included |
| Current acute decompensated heart failure | Not included | Not included | Not included |
| Heart failure due to active myocarditis | Not included | Not included | Not included |
| Symptomatic bradycardia | Not included | Not included | Not included |
| Rapidly progressing renal disease at time of randomization | Not included | Not included | Not included |
| **Confounder** | **Full Emulation Criteria Classification** | **Optimized Criteria Classification** | **Optimized Criteria Classification** |
| Type 2 diabetes at baseline | Confounder | Strict Inclusion | Confounder |
| MRA at baseline | Confounder | Strict Exclusion | Strict Inclusion |
| Atrial fibrillation or flutter on enrollment ECG | Confounder | Confounder | Strict Inclusion |
| Body-mass index | Confounder | Confounder | Confounder |
| Baseline eGFR (ml/min/1.73$m^2$) | Confounder | Confounder | Confounder |
| Age | Confounder | Confounder | Confounder |
| Sex | Confounder | Confounder | Confounder |
| Race | Confounder | Confounder | Confounder |
| LVEF | Confounder | Confounder | Confounder |
| NT-proBNP | Confounder | Confounder | Confounder |
| Hospitalization for heart failure | Confounder | Confounder | Confounder |
| Main cause of heart failure | Confounder | Confounder | Confounder |
| Dose of dapagliflozin | Confounder | Confounder | Confounder |

[*]Because the optimized cohorts were generated as OR-unions of multiple tree-derived rule paths, “Strict Inclusion” and “Strict Exclusion” indicate that the criterion was used as a strict component in at least one selected optimized rule path. These labels should not be interpreted as criteria required uniformly across all patients in the optimized cohort. Criteria not selected in the final optimized rule but retained in the propensity-score or Cox adjustment models were classified as confounders.

(a) Eligibility Criteria Subgroup: Beneficial

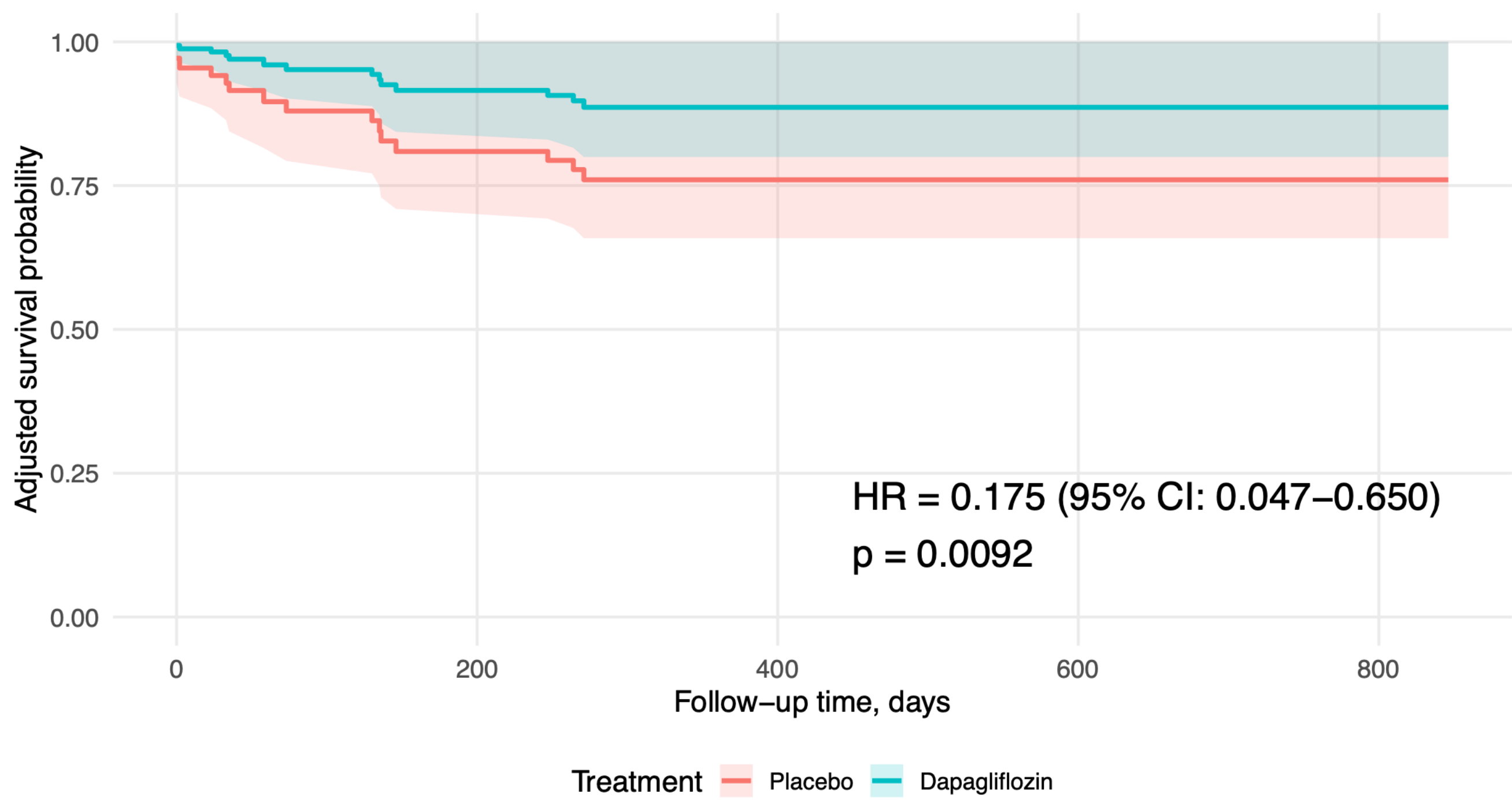

(b) Eligibility Criteria Subgroup: Harmful

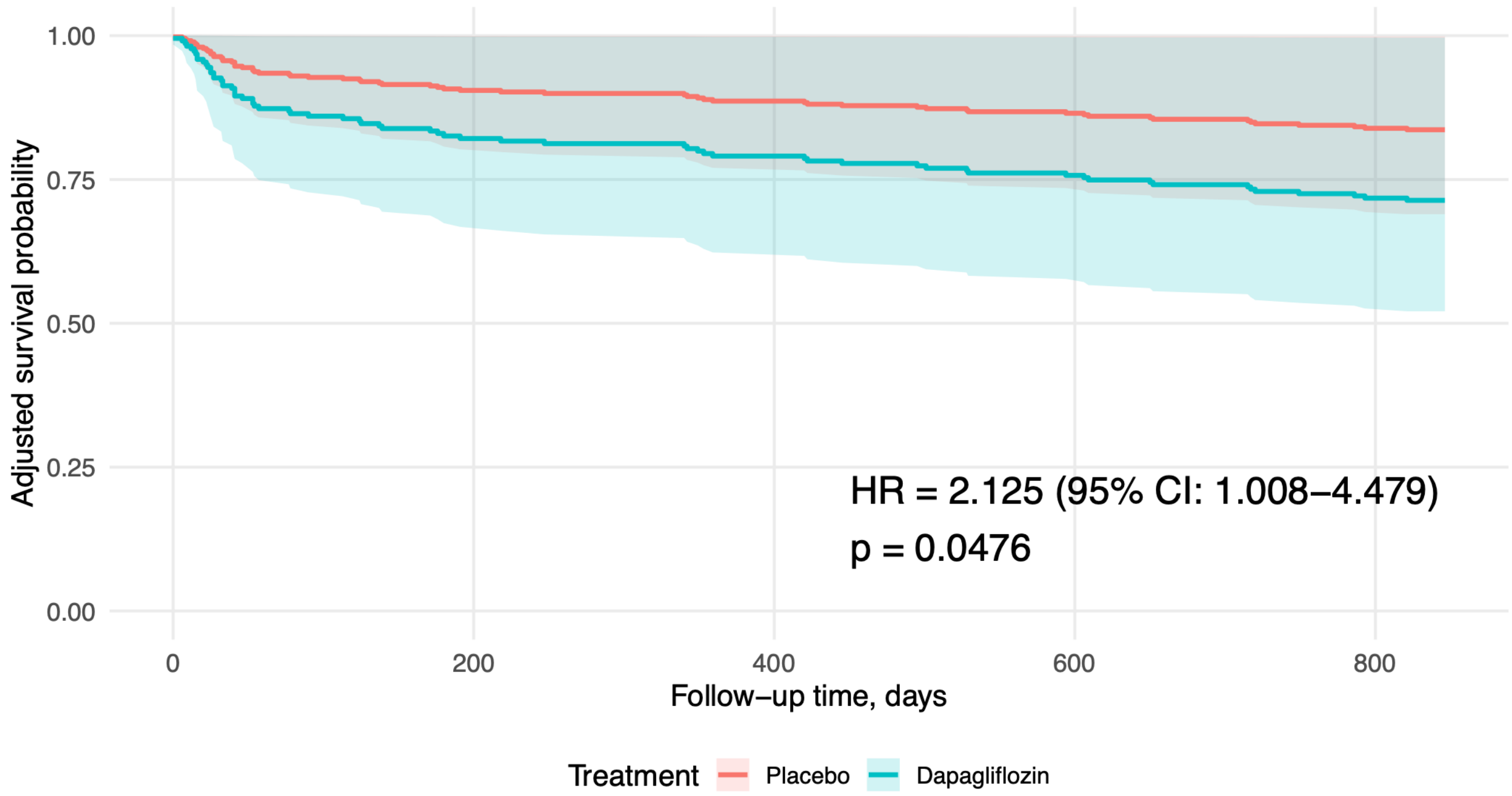

**Figure 4.** Survival Curve for the HTE-Informed Decision-Tree Eligibility Criteria Subgroup. Panel (a) shows the beneficial eligibility criteria subgroup, where dapagliflozin was associated with improved adjusted survival compared with placebo. Panel (b) shows the harmful eligibility criteria subgroup, where dapagliflozin was associated with worse adjusted survival compared with placebo. The red curve represents placebo, and the blue curve represents dapagliflozin; shaded areas indicate 95% confidence intervals. These results demonstrate that the HTE-informed decision-tree optimization identified clinically distinct subgroups with opposite treatment-effect directions.

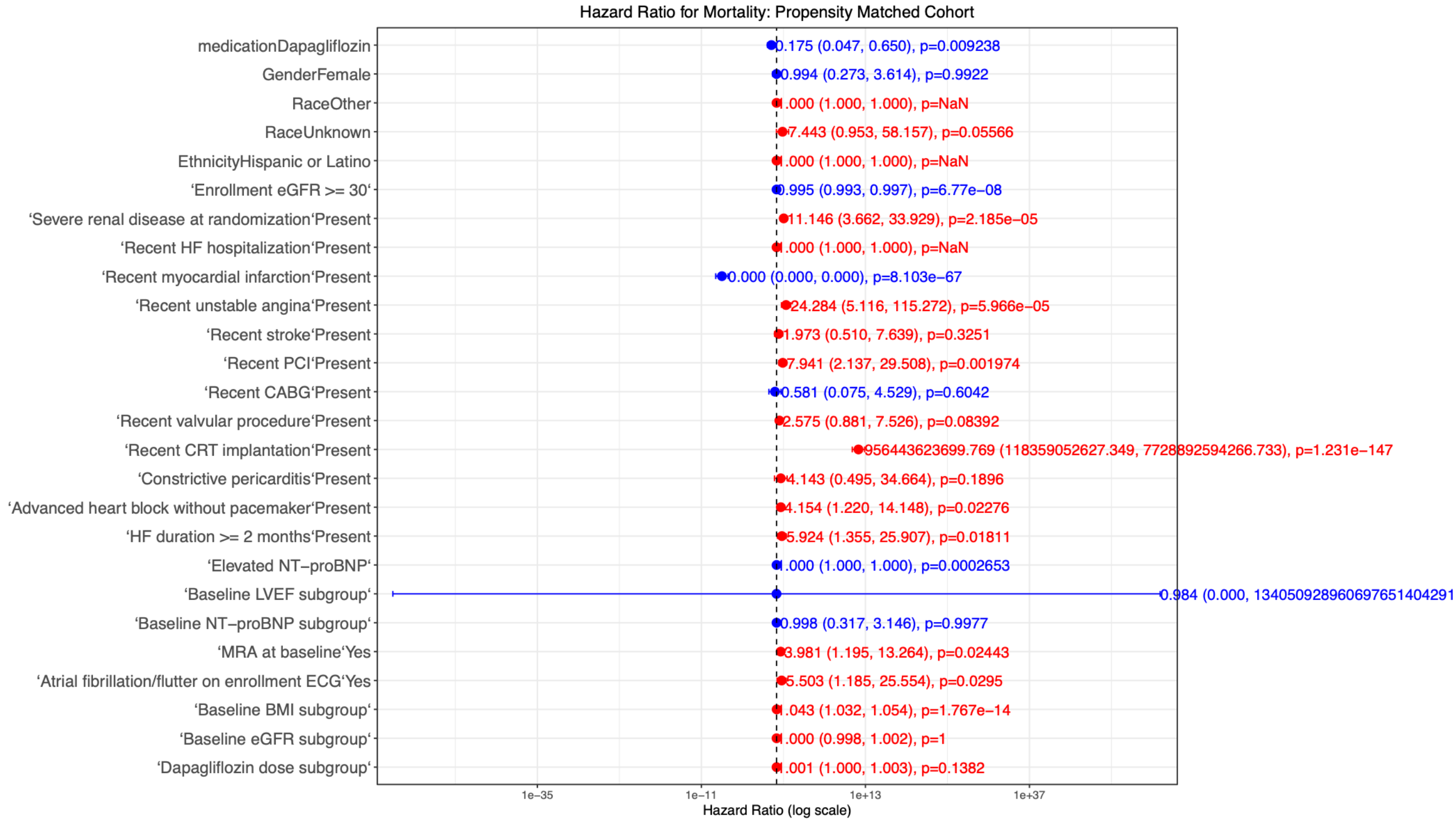

**Figure 5.** Hazard Ratios for Mortality in the HTE-Informed Decision-Tree Eligibility Criteria Subgroup (Beneficial). This forest plot presents hazard ratio estimates and 95% confidence intervals for dapagliflozin versus placebo and associated clinical covariates within the subgroup defined by the HTE-Informed decision-tree eligibility criteria optimization approach. Estimates were obtained from a Cox proportional hazards model fitted in the propensity score–matched cohort.

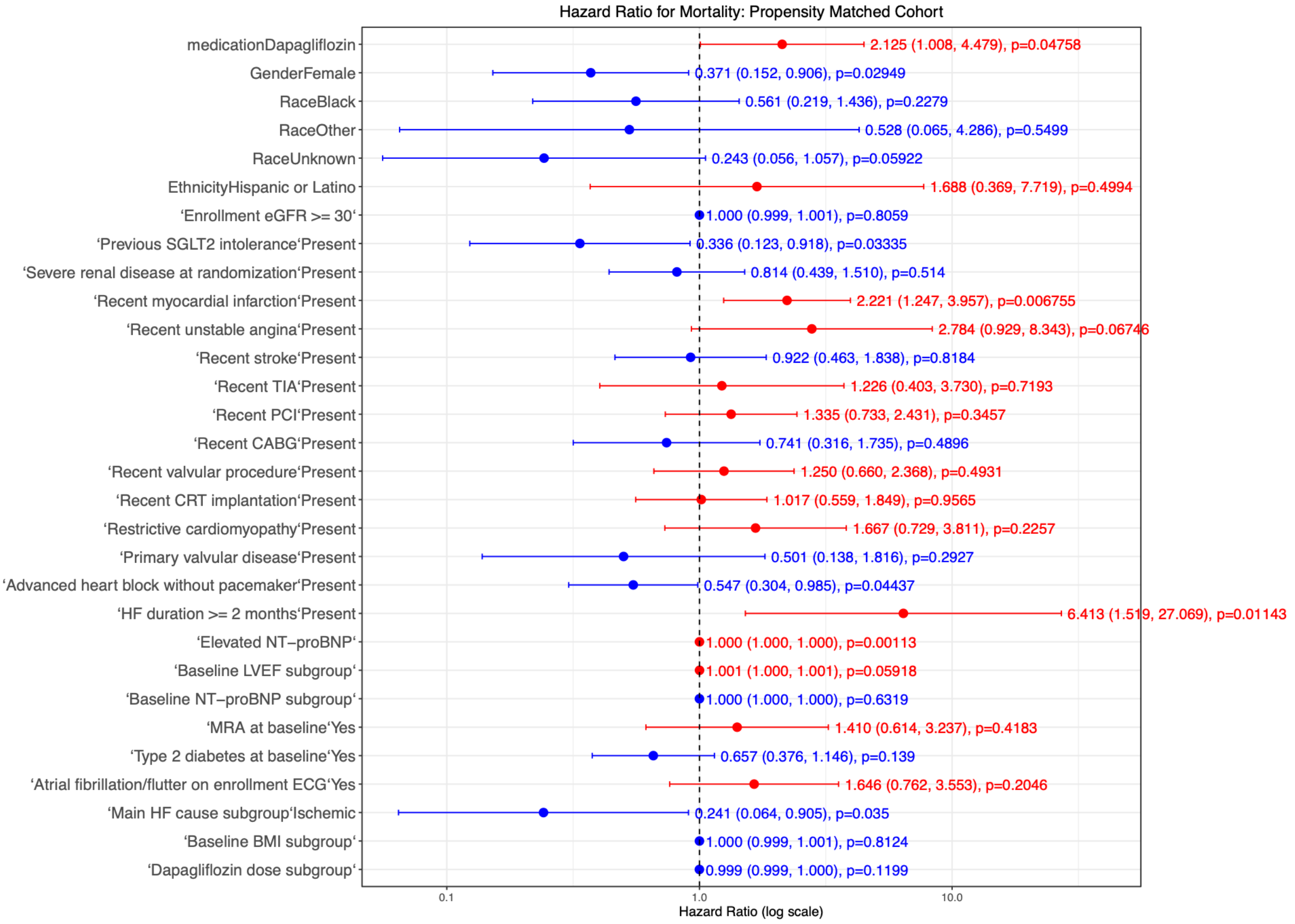

**Figure 6.** Hazard Ratios for Mortality in the HTE-Informed Decision-Tree Eligibility Criteria Subgroup (Harmful). This forest plot presents hazard ratio estimates and 95% confidence intervals for dapagliflozin versus placebo and associated clinical covariates within the subgroup defined by the HTE-Informed decision-tree eligibility criteria optimization approach. Estimates were obtained from a Cox proportional hazards model fitted in the propensity score–matched cohort.

## 3. Discussion

This study emulated the completed DAPA-HF trial using clinical data from the MCC. Analysis meeting DAPA-HF eligibility criteria showed no significant difference in mortality between placebo and dapagliflozin among patients. However, when individual-level heterogeneous treatment effects were estimated using machine learning and patients were stratified accordingly, a distinct subgroup emerged in which dapagliflozin was associated with a significantly higher risk of mortality. These findings suggest that HTE-guided stratification can uncover clinically meaningful treatment heterogeneity that is masked in average treatment comparisons.

However, our approach has several limitations. First, the current study focused only on a single HTE-based subgrouping strategy, which limits the scope of our findings. While splitting patients into one set of responder and non-responder groups is a logical first step, it may obscure more complex patterns of treatment heterogeneity. Exploring alternative combinations or interactions of subgroups, for example, comparing groups 2 and 3 or dynamically merging smaller strata, could reveal additional insights and provide a richer picture of treatment response. However, increasing the number of subgroups also introduces methodological risks. Smaller strata may become unstable, and repeated subgroup exploration increases the likelihood of false-positive findings. Prospective evaluation and external validation across other data sources are necessary to determine whether these HTE-derived partitions truly define groups that should be the focus of future RCTs.

Second, while we employed advanced HTE estimation technique (Meta-S) individually, we did not systematically compare them to a wider set of models. Each estimator has distinct assumptions and inductive biases, and subgroup definitions could shift depending on the chosen method. Relying on only two models may therefore overstate confidence in the findings. Future work should benchmark multiple HTE learners, including causal forests and ensemble-based approaches, to evaluate consistency and reduce model-specific biases. Ensemble strategies such as super learners or stacked estimators could also help stabilize treatment effect estimates by leveraging the strengths of multiple algorithms. Beyond improving predictive accuracy, such comparisons would highlight regions where models disagree, signalling patient subgroups where evidence is weak and where trialists should be especially cautious in interpreting HTE outputs.

Third, a key limitation of this study is that our method was evaluated on only one trial, which may limit the generalizability of the findings and introduce bias specific to this cohort. The results, particularly within the overlapping groups, could therefore reflect characteristics unique to this dataset rather than broadly applicable treatment effects. To strengthen the clinical relevance of our approach, we plan to discuss these findings with physicians to assess their plausibility and contextualize them against established medical knowledge.

## 4. Methods

### 4.1. Study Design and Trial Emulation Overview

Study design and causal validity considerations have already been emphasized in our prior MCC trial emulation work [20], and we therefore do not repeat them in detail here. Figure 7 illustrates the methodological framework for emulating the DAPA-HF trial using real-world data from the Mayo Clinic Cloud (MCC). The original DAPA-HF trial compared dapagliflozin and placebo in patients with heart failure and reduced ejection fraction (EF ≤ 35%) and found no statistically significant difference between the two treatments. To evaluate how these findings translate to real-world settings, we constructed an EHR-based emulation by identifying eligible patients from the MCC database in accordance with the trial's inclusion and exclusion criteria. For the dapagliflozin group, we included patients who met the Guideline-Directed Medical Therapy (GDMT) criteria [21] for heart failure and received dapagliflozin. The index time ($T_0$) was defined as the earliest time point at which dapagliflozin was prescribed within an eligible time segment. For the placebo group, we included patients who met the same GDMT criteria but did not receive dapagliflozin, and defined $T_0$ as the latest GDMT medication time within an eligible time segment.

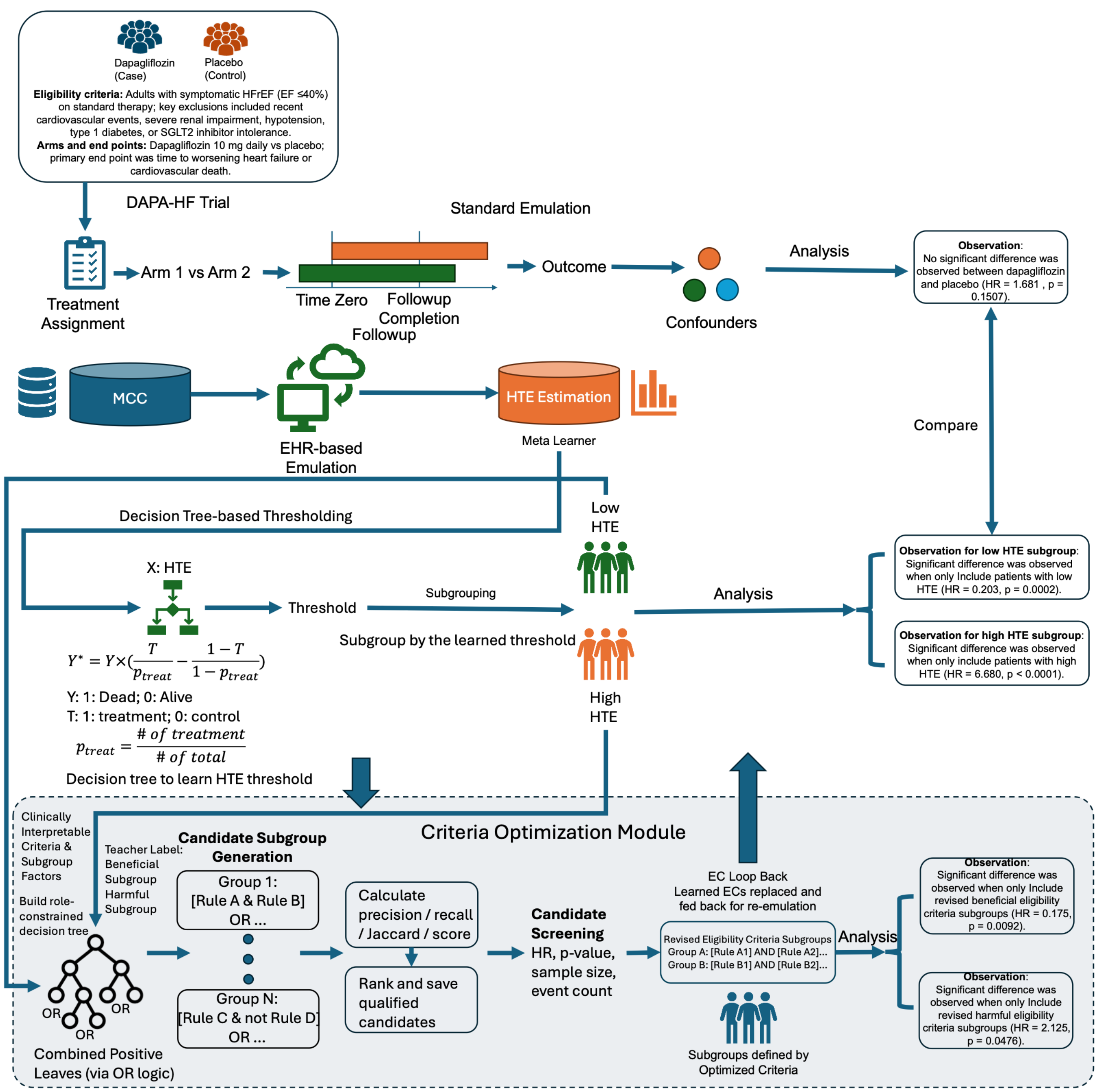

**Figure 7.** Overview of the Proposed Framework for DAPA-HF Trial Emulation, HTE-Guided Subgroup Discovery, and Eligibility Criteria Optimization Based on HTE-Informed Decision-Tree. This figure summarizes the workflow for emulating the DAPA-HF trial using EHR data from the Mayo Clinic Platform and applying heterogeneous treatment effect (HTE) modeling to optimize treatment comparisons. In the standard emulation, patients are selected using trial-inspired eligibility criteria, assigned to dapagliflozin or placebo arms, followed from the defined time zero to outcome or censoring, adjusted for clinical confounders, and evaluated using survival analysis. This standard emulation shows no significant difference between dapagliflozin and placebo in the overall cohort (HR = 1.681, p = 0.1507). To identify treatment heterogeneity that may be masked in the aggregate analysis, two HTE-based subgrouping strategies are applied. First, we use decision tree–based HTE selection to identify low- and harmful subgroups, revealing a significant protective association in the beneficial subgroup (HR = 0.203, p = 0.0002) and a significant adverse association in the harmful subgroup (HR = 6.680, p < 0.0001). Building on the standard trial emulation and HTE estimation results, the proposed framework integrates a criteria optimization module to translate HTE-derived treatment-response patterns into interpretable eligibility-criteria definitions. Clinically meaningful eligibility-related criteria, baseline clinical characteristics, and prespecified subgroup factors were used to construct rule-based candidate subgroups from HTE-enriched clinical paths. Multiple positive tree-derived paths were combined using OR logic, and the resulting candidate

subgroups were ranked using precision, recall, Jaccard index, sample size, treatment/control balance, and event availability before Cox-based screening. The final optimized beneficial and harmful criteria were then incorporated into the cohort-construction process, replacing the prior eligibility definitions, and the full emulation workflow was repeated using the revised cohorts. This process identified an optimized beneficial subgroup with stronger dapagliflozin-associated survival benefit (HR = 0.175, p = 0.0092) and an optimized harmful subgroup with increased mortality risk associated with dapagliflozin (HR = 2.125, p = 0.0476). Overall, the framework demonstrates how HTE-guided subgroup discovery and eligibility-criteria optimization can convert hidden treatment-effect heterogeneity into clinically interpretable revised criteria for identifying patients with more favorable or unfavorable treatment-response patterns.

## 4.2. Cohort Construction and Outcome Definition

Patient cohorts were constructed using data from the Mayo Clinic Cloud by applying the key inclusion and exclusion criteria consistent with the DAPA-HF trial. Additional adjustments were made to ensure demographic and clinical balance between individuals prescribed dapagliflozin and those receiving placebo. The primary outcome assessed was all-cause mortality.

## 4.3. Heterogeneous Treatment Effect (HTE) Estimation

We then estimated individual-level HTEs using approaches such as Meta Learner. To estimate individual-level HTEs, we employed one independent modelling strategies: the Meta-S Learner. This method was applied to generate patient-specific treatment effect estimates. The Meta-S learner (S-learner) is a single-model framework for estimating heterogeneous treatment effects. In this approach, treatment assignment is incorporated as an additional covariate in a unified predictive model, which learns the outcome as a function of both baseline covariates and treatment status. Individualized treatment effects are then estimated by computing the difference between predicted potential outcomes under treatment and control for the same patient. The S-learner is model-agnostic and can be implemented using flexible machine learning algorithms such as gradient boosting or neural networks. It is particularly suitable when treatment and covariates interact in complex, nonlinear ways, and has been widely adopted in causal machine learning literature [17].

## 4.4. HTE-Based Subgroup Stratification

Based on the estimated HTE scores, patients were stratified into subgroups using auto-learned quantile-based thresholds, while accounting for treatment assignment. Patients with HTE scores below the threshold were classified as likely to benefit from treatment (i.e., associated with lower mortality risk), whereas those above the threshold were considered less likely to benefit or potentially harmed (i.e., associated with higher mortality risk). To determine the HTE threshold, we used a shallow decision-tree regression model trained on the transformed outcome $Y^* = Y \times (\frac{T}{p_{treat}} - \frac{1-T}{1-p_{treat}})$, where (Y) represents the mortality outcome, (T) indicates treatment assignment, and ($p_{treat} = \frac{\#\ of\ treatment}{\#\ of\ total}$) is the overall treatment prevalence. This transformed outcome [22, 23] allowed the thresholding process to incorporate both mortality events and treatment/control assignment, rather than relying on mortality alone. The derivation showing why this transformed outcome can be interpreted as a treatment-effect pseudo-outcome is provided in Supplementary Note 1. Subgroup assignment incorporated both the predicted treatment effect and the therapy actually received, enabling identification of concordant (treatment aligned with predicted benefit) and discordant cases.

We explored a complementary strategy for subgroup identification. A decision tree model was trained using mortality as the outcome to learn an optimal threshold on the HTE score. This data-driven threshold was subsequently used to define beneficial (low HTE) and harmful (high HTE) subgroups.

## 4.5. Eligibility Criteria Optimization Based on HTE-Informed Decision-Tree

To translate the HTE-derived subgroup findings into interpretable eligibility criteria, we developed an eligibility-criteria optimization procedure based on HTE-informed decision-tree rule discovery. First, patient-level HTE estimates were used to define potential beneficial and harmful treatment-response regions. Patients below the learned HTE threshold were considered potential benefit-enriched patients, whereas patients above the threshold were considered potential harm-enriched patients.

Using these HTE-defined benefit and harm targets as reference labels, we then trained decision-tree–based rule models to identify clinically interpretable baseline criteria that approximated the HTE-enriched subgroups. Candidate predictors were restricted to baseline eligibility-related criteria and clinically meaningful subgroup factors, including demographic characteristics, renal function measures, cardiovascular history, baseline medication use, and prespecified clinical subgroup variables. Variables related to treatment assignment, survival outcomes, follow-up duration, and HTE scores were excluded from the rule-generation inputs to prevent information leakage and ensure that the final optimized criteria were based only on clinically available baseline information.

Terminal leaves and tree-derived rule paths were treated as candidate eligibility-criteria subgroups. Each rule path represented an interpretable combination of baseline clinical conditions. Rather than relying on a single terminal leaf, multiple positive leaves enriched for the HTE-defined benefit or harm target were combined using OR logic to construct candidate optimized cohorts; therefore, patients satisfying any selected rule path were included in the corresponding candidate cohort. Candidate cohorts were ranked and filtered using precision, recall, Jaccard index, cohort size, treatment/control balance, and event availability. Qualified candidates were then evaluated using the same downstream analysis framework, including propensity-score matching and adjusted Cox proportional-hazards modeling. Final beneficial candidates were required to have $HR < 1$ and ($p < 0.05$), whereas final harmful candidates were required to have $HR > 1$ and ($p < 0.05$).

After the final optimized beneficial and harmful criteria were selected, these criteria were incorporated into the cohort-construction step to generate revised emulated cohorts. The full emulation workflow was then repeated on the revised cohorts, including treatment assignment definition, follow-up construction, propensity-score matching, adjusted Cox modeling, and adjusted survival-curve estimation. This process allowed HTE-derived subgroup findings to inform eligibility-criteria refinement while preserving interpretability, clinical grounding, and consistency with the original trial-emulation analysis framework.

## Acknowledgments

This work is supported by a grant from the National Institute of Health (NIH) NIGMS (R00GM135488), NIH R01AG084236, and Mayo Clinic Department Chair funding. This work was supported in part by the Nancy Peretsman and Robert Scully Chair of Artificial Intelligence and Informatics at Mayo Clinic.

## Contributions

NZ and CT conceptualized and supervised the whole study. NZ led and managed this project. XL contributed to data processing, emulation design, and experiment. ML helped improve the experimental visualization. PL, XKL, JJ, and PP provided the domain expertise for the conduction of the emulation.  NZ and XL composed the initial draft of the report, and all authors reviewed the manuscript. The final manuscript was approved by all authors.

## Data and Code Availability

This study involves analysis of de-identified Electronic Health Record (EHR) data via Mayo Clinic Cloud. Data shown and reported in this manuscript has been extracted from the EHR using an established protocol for data extraction, aimed at preserving patient privacy. The data has been determined to be de-identified pursuant to an expert’s evaluation, in accordance with the HIPAA Privacy Rule. Any data beyond what is reported in the manuscript, including but not limited to the raw EHR data, cannot be shared or released due to the parameters of the expert determination to maintain the data de-identification. Contact corresponding authors for additional details regarding Mayo Clinic Cloud.

## Ethics and Consent to Participate

This study used retrospective electronic health record data from Mayo Clinic. The study did not involve prospective recruitment, intervention, or direct interaction with human participants. Data use was reviewed and approved by the Mayo Clinic Institutional Review Board (IRB) and was determined to be exempt research. The requirement for informed consent was waived by the Institutional Review Board because the study involved secondary analysis of existing clinical data and posed minimal risk to participants. All analyses were conducted in accordance with applicable institutional policies and privacy protections.